# Artificial Intelligence Models and Employee Lifecycle Management:

# A Systematic Literature Review

Saeed Nosratabadi[1], Roya Khayer Zahed[2,*], Vadim Vitalievich Ponkratov[3], Evgeniy Vyacheslavovich Kostyrin[4]

[1] Doctoral School of Economic and Regional Sciences, Hungarian University of Agriculture and Life Sciences, 2100 Godollo, Hungary, saeed.nosratabadi@gmail.com

[2] Department of Management, Faculty of Administrative Sciences and Economics, University of Isfahan, Isfahan, Iran, R.khayer@yahoo.com

[3] Department of Public Finance, Financial University under the Government of the Russian Federation, Moscow, Russian Federation, ponkratovvadim@yandex.ru

[4] Department of Finances, Bauman Moscow State Technical University, Moscow, Russian Federation, evgeniy.kostyrin@yandex.ru

* Corresponding Author: R.khayer@yahoo.com

Declarations of interest: none'

**Abstract**

**Background/Purpose:** The use of artificial intelligence (AI) models for data-driven decision-making in different stages of employee lifecycle (EL) management is increasing. However, there is no comprehensive study that addresses contributions of AI in EL management. Therefore, the main goal of this study was to address this theoretical gap and determine the contribution of AI models to EL.

**Methods:** This study applied the PRISMA method, a systematic literature review model, to ensure that the maximum number of publications related to the subject can be accessed. The output of the PRISMA model led to the identification of 23 related articles, and the findings of this study were presented based on the analysis of these articles.

**Results:** The findings revealed that AL algorithms were used in all stages of EL management (i.e., recruitment, on-boarding, employability and benefits, retention, and off-boarding). It was also disclosed that Random Forest, Support Vector Machines, Adaptive Boosting, Decision Tree, and Artificial Neural Network algorithms outperform other algorithms and were the most used in the literature.

**Conclusion:** Although the use of AI models in solving EL problems is increasing, research on this topic is still in its infancy stage, and more research on this topic is necessary.

**Keywords:** Artificial intelligence, deep learning, machine learning, human resource management, employee lifecycle, PRISMA, systematic literature review

## 1. Introduction

Innovations, new technologies, and the Covid 19 pandemic have posed new challenges to human resource management (HRM). These changes have not only required a new set of skills for employees, but also affected the way tasks are performed, and have intensified the platform economy and the emergence of platform workforces (Illéssy, Huszár, & Makó, 2021; Makó & Illéssy, 2020). In addition, information systems have greatly facilitated the processes of storing and collecting data related to individuals, which provide the basis for decision-making about the organization's workforce. Many statistical models are proposed in the literature for analyzing this information, but with the prevalence of artificial intelligence (AI) models, the use of these models in HRM has become common. Two features of AI models distinguish them from statistical models and have made the use of these models more popular than statistical models. Their first feature is the high performance of these models in nonlinear and noisy data (Ardabili et al., 2019; Nosratabadi, Szell, et al., 2020). The second feature is that these models have the ability to learn from the data to improve their performance. In other words, machine learning and deep learning models, which are subsets of AI models, are able to identify trends in data, even nonlinear and noisy data, during the training phase to classify the data or predict the behavior of phenomena based on identified patterns (Nosratabadi, Ardabili, Lakner, Mako, & Mosavi, 2021; Nosratabadi et al., 2020). Therefore, the AI models have been used to take advantage of these features and to find appropriate solutions to problems in different stages of HRM. However, there is no integrated and comprehensive study in the literature that identifies which HRM problems are addressed by AI models. Therefore, the present study was conducted to bridge this gap in the literature using a systematic review study to determine how AI has been able to help HR managers. In order to evaluate the contribution of AI in HRM, the present study uses the employee lifecycle (EL) model. The EL model is actually an HRM model that explains all the different life stages of the workforce from the time they are hired to the time they leave the organization. Inspired by this model, the present study aims to identify contributions of AI models to each stage of the EL management. Therefore, the research questions that the present study intends to answer are:

What AI models have been used in each stage of EL?

What human resource problems have AI models been used to solve?

What data sources have been used to test AI models?

This research contributes to the HRM literature and AI literature. The findings of this study provide HR managers with suggestions for selecting the appropriate AI model to address issues related to each stage of HRM. In the continuation of this article, first the methodology used in this article is described in detail and then the findings of this article are presented in order to answer the research questions, which is accompanied by a discussion on the findings and the conclusion.

## 2. Literature Review

Machine learning is driving an explosion in AI capability, helping software make sense of the messy and unpredictable real world. Today, machine learning is used in various sciences and can examine a large amount of data and discover certain trends and patterns that are unknown to humans. On the other hand, with the help of machine learning, there is no need for humans to intervene directly in every step of the project process. Therefore, the machine can make predictions on its own and also improve its algorithms to increases accuracy and efficiency. These algorithms perform well in examining multidimensional and multivariate data in dynamic or unknown environments.

Today, AI algorithms are expanding widely in organizations. These algorithms can promote and improve the organization's performance Olan et al. (2022) and prevent problems in the organization. When a system breaks down, it imposes huge costs such as time, productivity, and money to the business, machine learning and deep learning algorithms make it possible to quickly identify the cause of a problem and also predict the problems and solve them. On the other hand, with artificial intelligence algorithms, suitable data can be prepared for detailed analysis on important business decision criteria. For example, the behavior of different groups of buyers can be deeply investigated and better offers can be made to them (Jiang et al., 2022) to increase customer satisfaction. In an online store, one of the metrics is the amount of time a customer spends on a particular product page and AI here provides advanced analytics. In addition, AI can help predict organizational resource planning. By searching among the collected data, AI can make predictions that increase the ability of the organization. It also identifies seasonality in the business and make recommendations on increasing or decreasing production accordingly. Another application of AL in organizations is that it helps organizations to identify their behavioral patterns by considering the history of customers and predict how much of what type of product they should produce in the future (Jiangang et al., 2022). On the other hand, with the help of AI, it is possible to simplify sales (Irfan, et al, 2022), accounting (Leitner-Hanetseder et al., 2021), inventory (Praveen, Farnaz, & Hatim, 2019) etc., and create a centralized platform for managing customer relations (Deb, Jain, & Deb, 2018), as well as the product and sales life cycle (Ren, Patrick Hui, & Jason Choi, 2018). One of the important considerations of AI in organizations is to identify new opportunities for sales and marketing. Artificial intelligence and machine learning allow a business to not only identify the buying behavior of customers (Jiang et al., 2022), but also investigate what each person is willing to buy. AI can identify processes that cause unacceptable energy consumption in the organization or are mechanically inefficient, thereby helping to reduce energy consumption and resource wastage (Giaglis, 2001).

Organizations face different challenges for HRM, whether it is during recruitment, or during the collaboration with the employee, or when the employee's relationship with the organization is to be terminated for any reason (Susmita and Singh, 2022). Human resource is the key asset of the organization, which is known today as organizational capital. Organizations seek to attract the best and most suitable people in the organization, carefully implement the socialization process and familiarize employees with the basic principles and the main culture of their organization, as well as the necessary training on how to do the job (Morozevich et al., 2022). These are to keep the capabilities of the employees up to date and prepare them to adapt to the work environment. On the other hand, compensation is one of the most important measures of HRM in creating motivation in employees. An effective compensation system can make the employee feel a sense of belonging to the organization and consider the organizational goals as part of their own goals. Therefore, employee lifecycle management is brought up and its importance becomes necessary. The EL model proposed by Peisl and Shah (2019) is an HRM model that explains the challenges of human resource management at different stages of a workforce's life from the time he/she is recruited to the time he/she leaves the organization. This model constitutes of five stages of recruitment, on-boarding, employability and benefits, retention, and off-boarding.

In this model, the recruitment phase includes all the processes that lead to the recruitment of a new employee. Recruitment is the process of searching, evaluating and bringing in new talents when a specific job position is vacant. This Process begins when an employee resigns, or a new position is created to meet the needs of the company. Therefore, recruitment is a need-based procedure that occurs only when there is an immediate need for it. This usually involves advertising a job vacancy and letting people know that the company is looking for a worker/employee with a particular talent or skill.  At on boarding stage, the staff is provided with the necessary information and tools to be more efficient and integrate into the culture of the organization (Khayer Zahed, Teimouri, & Barzoki, 2021). Onboarding or aligning new employees in the company is the process of adapting these people to the company's activity process as well as its organizational culture. It also includes providing the tools and information needed to increase the productivity of the workforce in the team. According to Peisl and Shah (2019), the next stage of an EL is employability and benefits. Employability refers the ability to retain the employee and, if necessary, "move" him/her to a new job and role in the same organization to meet new job needs, and benefits mean how to assign financial and non-financial benefits to employees in order to create a sense of belonging and commitment to the organization. Retention refers to the mechanisms by which HRM seeks to maintain the employees and their potentials. Employee retention depends on a combination of factors including flexible working conditions, professional development opportunities, and company culture, and more.  The last stage of an EL is called off-boarding, in which the employee, for various reasons such as finding a new job, retirement, dismissal, personal reasons, stops working with the organization.

3. **Methodology**

In this study, four criteria were set to ensure finding the maximum number of articles related to AI models used in HRM (see Figure 1). The first criterion is to use Preferred Reporting Items for Systematic Reviews and Meta-Analyses (PRISMA) method to obtain final articles. The PRISMA is an evidence-based systematic literature review method that consists of four stages (1) identification (2) screening, (3) eligibility, and (4) inclusion (Moher, Liberati, Tetzlaff, & Altman, 2010) to systematically maximize the possibility of finding the most relevant articles. The second criterion is the selection of databases in which the search for articles took place. For this purpose, two databases, i.e., Scopus and Web of Science, were used. Scopus covers 42,180 journals, conference proceeding, and book series and web of Science includes 21,894 journals, books, and conference proceedings while, the overlap rate of articles in these two databases is 99.11% (Singh, Singh, Karmakar, Leta, & Mayr, 2021). The third criterion for accessing the maximum number of related articles is the use of appropriate keywords. Table 1 summarizes the literature search strategy implemented in this study. It is worth mentioning the search inquiry took place among article title, abstract, keywords.

Table 1. Literature search strategy

| Query Terms |
| --- |
| • "human resource*" OR employee OR "human capital" OR staff OR HR OR HRM |
| AND |
| • artificial intelligence OR AI OR "machine Learning" Or "deep learning" |

The fourth criterion used in this study to validate the path to find articles ready for review is the use of a multidisciplinary approach when searching for articles. This approach allows the search for keywords in all journals of different disciplines and this search is not restricted to certain journals or specific categories of journals.

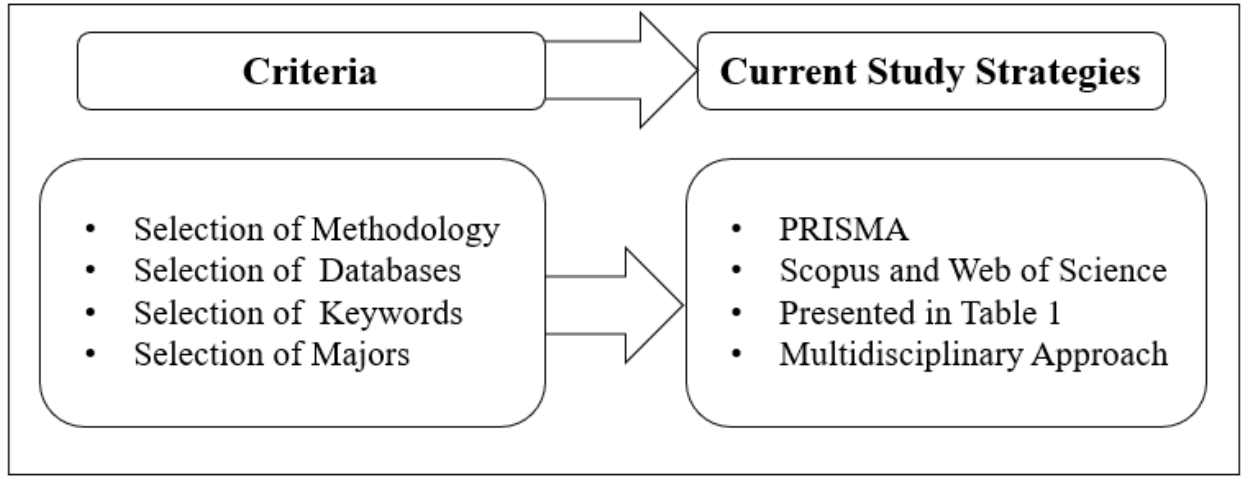

Figure 1. The current study strategies to validate the findings

The present study names the articles that were eventually identified for the review as the study database. As mentioned above, the PRISMA model was used to form the study database. Figure 2 summarizes the PRISMA model steps performed in this study. In the identification step, the keywords presented in Table 1 were searched in Scopus and Web of Science databases. It should be noted that this search was conducted in August 2021 and no time limit was considered for it. However, the search for articles was limited to original English-language articles in journals or conference proceedings. This means that in this search, articles written in another language, review articles, as well as documents that were either a book or a chapter of a book were removed. The output of the first stage resulted in the identification of 6753 articles. In the screening phase, a copy of duplicate articles found in both databases was removed. The output of this step was to identify 6050 unique articles. The second screening step examined the titles and abstracts of the papers and excluded those that were irrelevant. At this point, the criteria for keeping relevant articles were that they should use an AI model to address a problem in HRM. This step resulted in the identification of 580 relevant articles. Eligibility is the third step of the PRISMA and the whole text of the output articles from the screening step is thoroughly examined in this step, and only those articles that are absolutely relevant to the research's aim advance to the inclusion phase. Following a thorough examination of the complete text of 580 articles, 23 relevant articles were identified and advanced to the inclusion stage. The inclusion step resulted in the creation of the study database, which now contains 23 papers and is ready for further analysis.

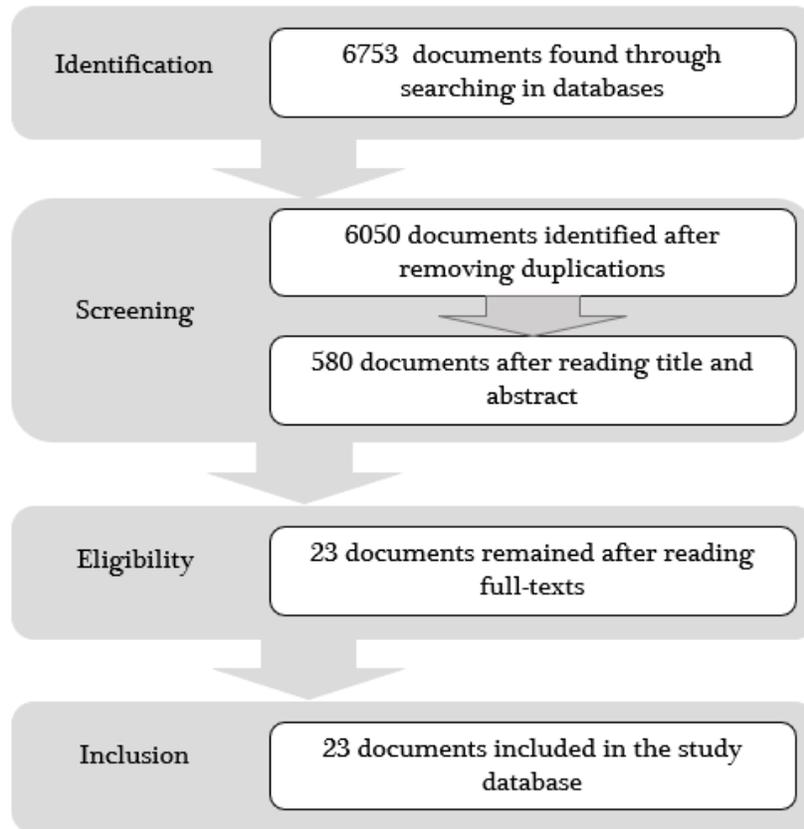

Figure 2. Systematic selection of the study database using PRISMA

## 4. Findings and Discussion

Figure 3 depicts a trend of publications using AI models to solve a problem related to HRM. Findings disclosed that the use of AI models in the field of HRM is very new as the first article was published in 2014. The present study found 23 related articles in the literature, of which 14 were journal articles and 9 were conference papers.

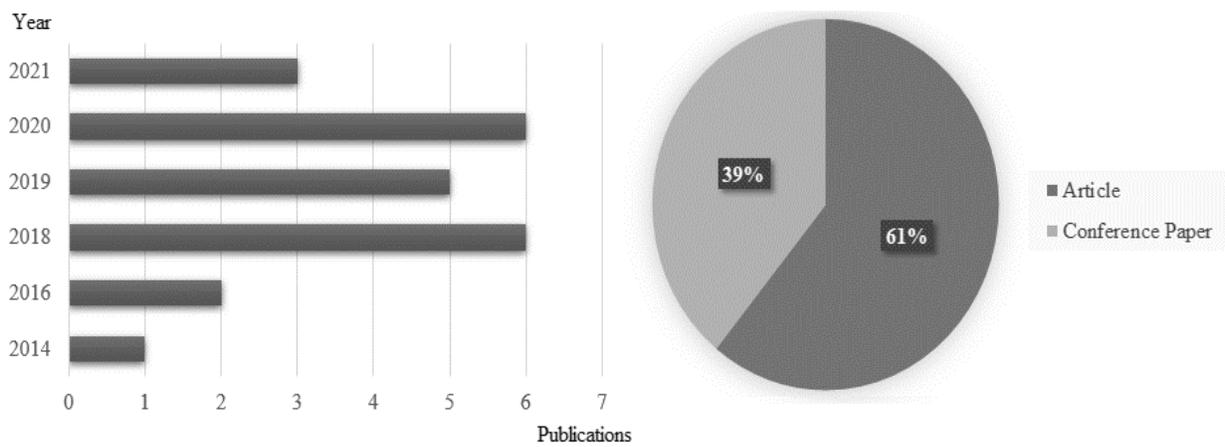

Figure 3. Categorizing the reviewed articles by year of publication and document types

21 unique sources have published the articles (i.e., 14 journals and 9 conference proceedings) from which Journal of 'Automation in Construction' and 'ACM International Conference Proceeding Series' in 2018 have published 2 articles on this topic. The full list of these journals and conference proceedings is given in Table A1 in the Appendix.

The articles found by the PRISMA method were analyzed from three aspects, and their findings are presented in the form of three subsections below. First, it was determined what problems the AI models were used to solve. In the next subsection, the articles were analyzed in terms of data sources. That is, it revealed what kind of HR data AI algorithms have been implemented on. Finally, AI algorithms that have been used in different phases of the EL management have been reviewed and the models whose performance has been repeatedly confirmed in the literature have been explained.

4.1.   Applications of Machine learning models in Human Resource Management

Once an employee arrives, he or she embarks on an adventurous journey and goes through various stages in his/her employment lifecycle. HRM experiences different challenges at each stage of the EL. Table 2 categorizes the reviewed articles based on the problem they addressed at each stage of the EL. Although the distribution of articles is almost the same at different stages, the share of articles dealing with employee attrition and off-boarding has been higher.

Table 2. Categorizing the reviewed articles by their solution in employee lifecycle

| | Employee Lifecycle | | | | |
|---|---|---|---|---|---|
| | **Recruitment** | **On-Boarding** | **Employability and Benefits** | **Retention** | **Off-Boarding** |
| Sources | Chuang, Hu, Liou, and Tzeng, (2020), Xie (2020), Zaman, Kamal, Mohamed, Ahmad, and Zamri (2018), N. Li, Kong, Ma, Gong, and Huai (2016) | X. Li et al. (2021), Kaewwiset, Temdee, and Yooyativong (2021), Liu, Li, Wang, and He (2019), Akhavian and Behzadan (2016), Colomo-Palacios, González-Carrasco, López-Cuadrado, Trigo, and Varajao (2014) | Singer and Cohen (2020), Jayadi, Jayadi, and Firmantyo (2019), Liu, Wang, et al. (2019), Long, Liu, Fang, Wang, and Jiang (2018) | Zhe and Keikhosrokiani (2021), Moyo, Doan, Yun, and Tshuma (2018), Jebelli, Khalili, Hwang, and Lee (2018) | Jain, Jain, and Pamula (2020), Fallucchi, Coladangelo, Giuliano, and William De Luca (2020), Anh et al. (2020), Khera and Divya (2018), Yadav, Jain, and Singh (2018), Liu et al. (2018), Zhao, Hryniewicki, Cheng, Fu, and Zhu (2018) |

*4.1.1.   Recruitment*

The most important challenge of the recruitment phase is identifying and selecting the right person who, firstly, has the most skill matching with the job and, secondly, has the necessary commitment to continue working. Recruitment, training and retaining an employee is very costly and employees become the intellectual capital of the organization and therefore, replacing them will be very difficult and costly for the organizations. Hence, Chuang et al. (2020), N. Li et al. (2016), Xie (2020), and Zaman et al. (2018) have tried to provide models that optimize the process of recruiting and selecting the right person for the job. Chuang et al. (2020) consider recruitment as a multiple-attribute decision-making (MADM) problem and use a rough set theory (RST) model to optimize staff recruitment by prediction of person skill match. Because they believe that the previous models were based on human judgment and taste, but the use of MADM makes hiring employees more purposeful and objective. They use performance data from a Chinese food company to test the model. Zaman et al. (2018) applied a decision tree (DT) classification technique model to predict the degree to which job skills and staff skills match in order to recruit talent. To do so, they used the data available on users' LinkedIn to test the model. N. Li et al. (2016) develop a K-nearest neighbors (KNN) model to predict skill match of the potential candidates. They also proved the accuracy of their proposed model using UCI Machine Learning Repository data. Xie (2020) designs a hybrid machine learning model of Latent Factor Model-multi Grained Cascade Forest (LFM-gcForest) to measure the degree to which employees' skills and job required skills are matched. They used secondary data from Africa Health Placements to test the model. The findings of this study showed that the LFM-gcForest model plays an important role in the HR recruitment system in the intelligent manufacturing industry.

All these studies, in turn, have attempted to use AI models to measure the degree to which an individual's skills match the skills needed to work so that they can assist HRM in making decisions about hiring people. Hence the following proposition can be derived from these findings:

P1: AI models help HRM in job-person skills match prediction.

*4.1.2. On-Boarding*

The findings revealed that there are studies in the literature that have used machine learning and deep learning models to address problems in the on-boarding stage of EL. In general, studies related to this stage of EL have focused more on staff training. For instance, Kaewwiset et al. (2021) and Liu, Li et al. (2019) use random forest (RF) model to enhance and personalize staff training and to predict the potential growth of employees in the workplace. Using secondary data from the HRM information system database and interpersonal environment factors, Liu, Li, et al. (2019) develop a quantitative model that predicts an employee's growth rate at different stages of employment. Findings of this study showed that relationships with colleagues and the quality of relationships with people are very important and necessary for employee development. Using the Artificial Neural Network (ANN) model, Colomo-Palacios et al. (2014) also propose a model that can anticipate the competencies required by members of software development teams and subsequently suggest related development programs.

Besides training purposes, two other articles were found that developed quantitative models using machine learning models for the management of working people. In order to manage occupational health and safety, for example, X. Li et al. (2021) developed the Multi-task Cascaded Convolutional Networks (MTCNN)-MobileNet- Long Short-Term Memory (LSTM) model, which allows them to monitor workers' health and generate personalized safety and health alerts for workers in high-risk jobs. Akhavian and Behzadan (2016) also develop an ANN model that is able to track workers body movements by tracking their smartphones, to examine the behavior and the state of construction workers.

These studies disclose that the main purpose of using machine learning and deep learning models at this stage of the EL was to identify the skills needed by the person to design personalized training programs and predict employee growth rates. Therefore, the second proposition of this study is presented as follows.

P2: AI models help HRM in staff training planning.

### 4.1.3. Employability and Benefits

In the employability and benefits stage of EL, articles are categorized that focus more on issues related to staff promotion. Liu, Wang et al. (2019), for example, used the Adaptive Boosting (AdaBoost) model to examine the promotion and advancement of employees in the workplace, and tested this model in data from a state-owned enterprise in China. Long et al. (2018) also use the RF to develop a model for predicting employee promotion and use data from a Chinese state-owned enterprise to test the model. Findings of this study show that job experience (by year), number of positions held and the position level (i.e., seniority level) in the organization are the factors affecting staff promotion. Jayadi et al. (2019) develop a Naive Bayes (NB) classification method to predict employee performance. Utilizing the available data from 310 employees from the KAGGLE database in 2019, Jayadi et al. (2019) proved that their proposed model has a good power to predict employee performance. In addition to these studies, Singer and Cohen (2020) use the Classification and Regression Trees (CART) model to design a model for the prediction of the absence of employees and the service compensation system in order to prevent the absence of employees of a Brazilian company. These studies have used AI models to help HRM to predict employee performance and predict employee promotion. Therefore, the third proposition of this research is designed as follows.

P3: AI models help HRM in employee promotion prediction.

### 4.1.4. Retention

The articles categorized in the retention mainly focus on the issues of work quality and employee well-being. In other words, these articles consider the creating proper working conditions as employee retention requirements. Zhe and Keikhosrokiani (2021), for example, use the extreme learning adaptive neuro-fuzzy inference system (ELANFIS) model to develop a model for predicting the mental workload of knowledge workers (i.e., Delft University of Technology students). To manage the well-being of employees and their mental health, Jebelli et al. (2018) also uses an Electroencephalography (EEG) device (for the data collection) and the Support Vector Machine (SVM) model (for the data analysis) to design a model to predict the stress of construction workers. In addition, Moyo et al. (2018) use multinomial logistic Regression (MLR) model to develop a model for predicting the length of practice of employees in the health sector. In fact, they came up with a quantitative model that has the ability to accurately predict how long health care workers will stay in their jobs. A look at the objectives of these articles reveals that they try to predict the level of stress, mental and physical health of employees by using machine learning models so that they can use them in managing and improving work quality and employee well-being. Therefore, the fourth proposition of this research is written as follows.

P4: AI models help HRM in work quality management.

### 4.1.5. Off-Boarding

Predicting off-boarding and factors affecting employee attrition has been one of the most important trends in the HRM literature. There are studies that have tried to use machine learning models to predict employee attrition. Jain et al. (2020), for example, claim the RF model to be the best model for predicting employee attrition. Khera and Divya (2018) used the SVM model to predict employee attrition and tested the model by archiving employee data (including 1,650 employees) from three Indian IT companies. Fallucchi et al. (2020) use the Gaussian Naïve Bayes (GNB) classifier model to analyze how objective factors affect employee attrition and test the predictive accuracy of their model using IBM analytics data that includes 35 features and 1,500 samples. Yadav et al. (2018) compare the accuracy of RF, AdaBoost, DT, Logistic Regression (LR) and SVM models to create accurate and reliable models that optimize the cost of hiring and retaining quality staff. They report that DT outperforms other models in prediction of attrition in their dataset. The findings of this article showed that salaries and other financial aspects and promotions are not adequate incentives for retaining employees. On the other hand, since employee turnover costs are high in organizations, Zhao et al. (2018) using the Extreme Gradient Boost (XGBoost) model provide a model for predicting employee turnover. To analyze the model, they employed data from an IBM database and a bank database (9089 bank employees and 1470 IBM employees). Liu et al. (2018) also compare the accuracy of LR, Support Vector Classifier (SVC), RF, and Adaboost in prediction of employee turnover. Besides, Anh et al. (2020) combine SVM, LR and RF models to develop a model to predict future employee churn. They test the accuracy of their model on the data of 1470 employees of an organization. These findings show that AI models have the ability to predict employee attrition and employee turnover and HR managers can use these models to identify the effective causes of employee attrition so that they can avoid the high costs that the outflow of human capital imposes on the organization. Subsequently, the fifth proposition of this study is designed as follows.

P5: AI models help HRM in employee attrition prediction and employee turnover prediction.

*4.1.6. Contributions of AI Models to Employee Lifecycle Management*
The findings of this study illustrated that the use of AI predictive models in HRM is increasing, and these models provide managers with appropriate and approved tools through which they can cope with the challenges in each stage of EL. In this study, five propositions were designed based on the purposes for which an AI model was used. A summary of these proposals is given in Figure 4. The stage of recruiting and selecting the right person for a task is one of the most important stages of HRM and managers face many challenges to select the right person. According to the first proposition of this article, AI models can help managers in deciding to choose the right person for the task by predicting the degree to which the skills of the individual are matched with those skills needed for the job. Job and task dynamics, innovations, the use of new technologies, changes in organizational strategies are some of the factors that impose new requirements on employees to perform tasks. The second proposition of the present study clarified that at this stage, AI models have the ability to propose personalized training programs to develop the skills of individuals by examining personal skills and required skills. Creating motivation and a sense of commitment in employees is another important challenge of HRM. Therefore, managers design reward and promotion systems through which employees can perform their best performance in the organization. The third and fourth propositions of this research refer to these issues and state that AI models can both predict the performance and promotion of employees and can help managers in managing the quality of work. Up to this point in EL, organizations have spent a lot of money on recruitment, training, promotion and retention of the employees, and the employees are the intellectual capital of the organization and losing them will be very costly for the organizations. Hence, managers try to minimize and manage employee attrition. In this regard, the fifth proposition of this study shows that AI models can help HR managers in managing the attrition of employees by predicting the factors affecting employee attrition and employee turnover.

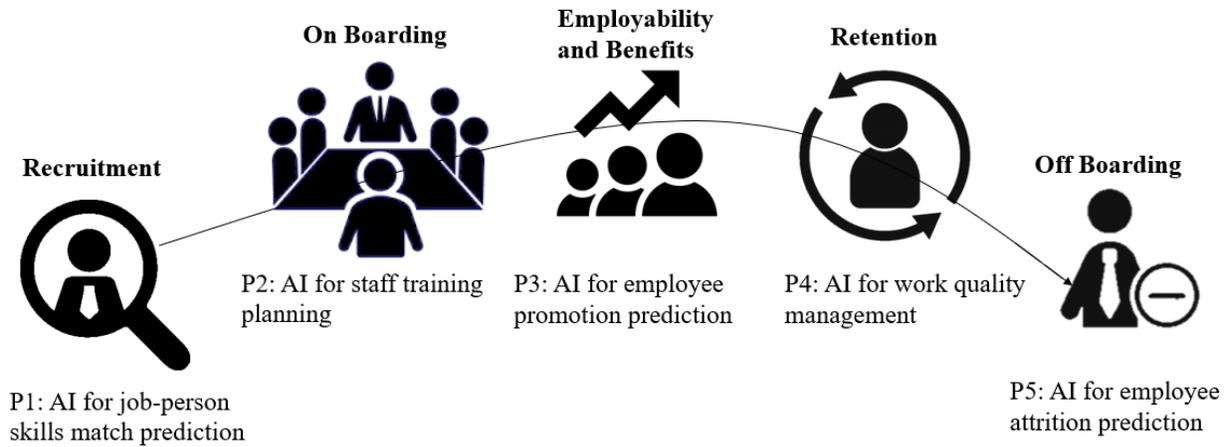

Figure 4. Contributions of AI models in different stages of employee lifecycle

4.2. Data Sources

One of the most important issues in implementing AI models is the data. In other words, it is very important where the data source comes from. It is revealed that the articles to test their proposed models used both primary data (collected through case studies, observations, user profiles on social media, and questionnaires) and secondary data (which have been collected from databases such as Kaggel.com, UCI Machine Learning Repository, IBM Analytics, and past records of a case study). Figure 5 shows the taxonomy diagrams of data sources.

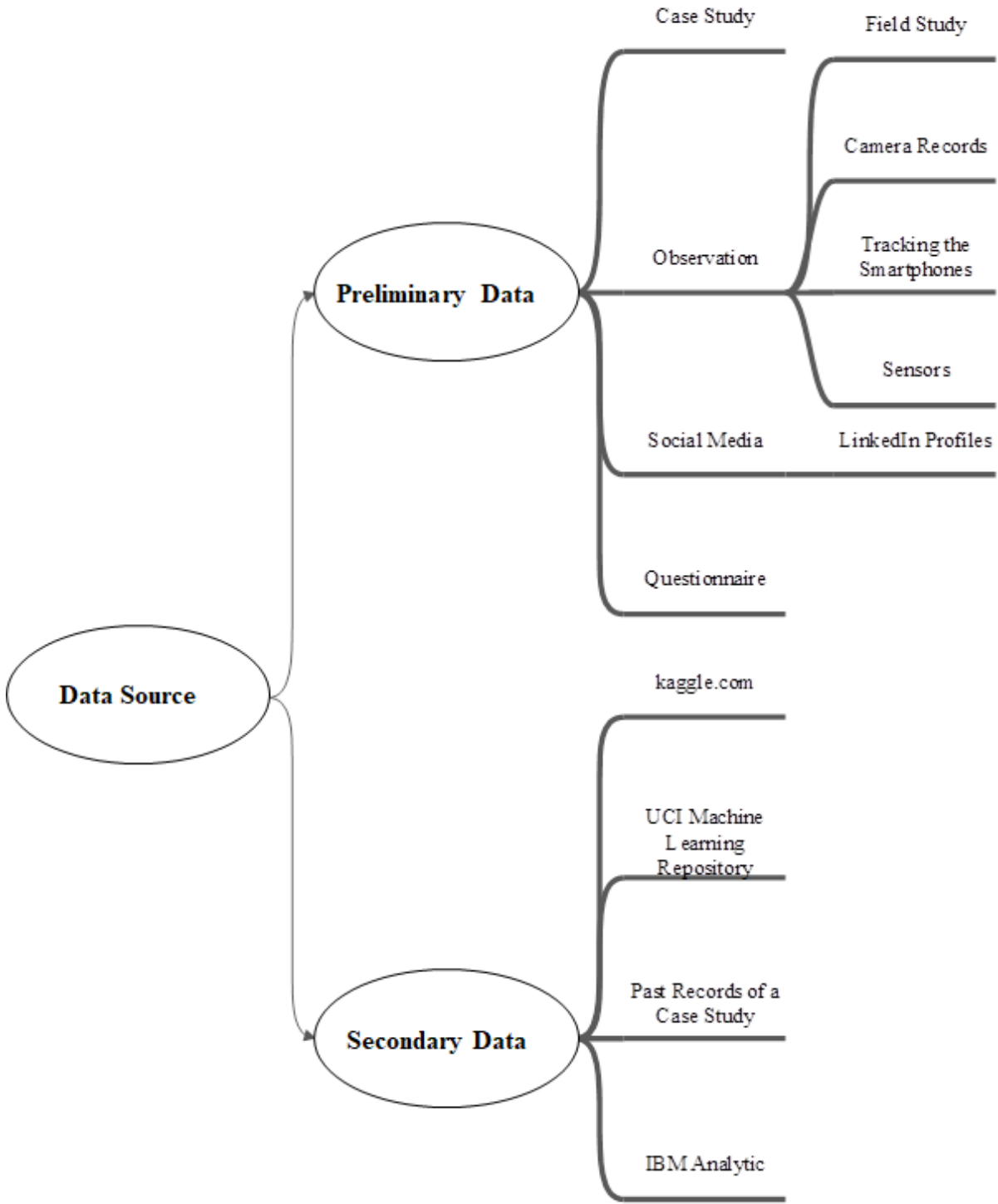

Figure 5. A taxonomy of data sources used for AI models to solve a human resource management problem

### 4.3. Artificial Intelligence Models

A closer look at the models used among the reviewed articles shows that these articles, in total, examined the performance of 26 different models (see Table 3). The procedure has been that each article first examines the performance of several models and then introduces the model that had the highest accuracy (or the lowest level of error) as the main model of that article. In this article, in the findings and discussion section, we only reported the model that had the highest performance. In the end, it was found that 16 of these models presented in Table 3 had the lowest level of error and their higher performance was repeated among different articles.

Table 3. Machine learning models used in human resource management literature

| Methods | Source | Methods | Source |
| --- | --- | --- | --- |
| AdaBoost | Liu, Wang, et al. (2019), Yadav, Jain, and Singh (2018), Liu et al. (2018) | LR | Fallucchi et al. (2020), Anh et al. (2020), Liu, Wang, et al. (2019), Liu, Li, et al. (2019), Yadav et al. (2018), Liu et al. (2018), Zhao et al. (2018), Akhavian and Behzadan (2016) |
| ANFIS | Zhe and Keikhosrokiani (2021) | LSVM | Fallucchi et al. (2020) |
| ANN | Zhao et al. (2018), Akhavian and Behzadan (2016), Colomo-Palacios, González-Carrasco, López-Cuadrado, Trigo, and Varajao (2014) | MLR | Moyo et al. (2018) |
| DT | Kaewwiset, Temdee, and Yooyativong (2021), Jain, Jain, and Pamula (2020), Fallucchi et al. (2020), Zaman et al. (2018), Moyo et al. (2018), Yadav et al. (2018), Zhao et al. (2018), Akhavian and Behzadan (2016) | MLP | Singer and Cohen (2020), Liu, Li, et al. (2019) |
| ELANFIS | Zhe and Keikhosrokiani, (2021) | MTCNN-MobileNet-LSTM | X. Li et al. (2021) |
| Fed-GRU | X. Li et al. (2021) | NB | Singer and Cohen (2020), Jayadi et al. (2019), Moyo et al. (2018), Zhao et al. (2018) |
| Fed-LSTM | X. Li et al. (2021) | NB-MB | Fallucchi et al. (2020) |
| Fed-SWP | X. Li et al. (2021) | Ordinal CART | Singer and Cohen (2020) |
| GBoost | Zhao et al. (2018) | RF | Kaewwiset et al. (2021), Singer and Cohen (2020), Jain et al. (2020), Anh et al. (2020), Liu, Wang, et al. (2019), Liu, Li, et al. (2019), Yadav et al. (2018), Long et al. (2018), Liu et al., (2018), Zhao et al. (2018) |
| GNB | Fallucchi et al. (2020) | RST | Chuang et al. (2020) |
| KNN | Singer and Cohen (2020), Fallucchi et al. (2020), Zhao et al. (2018), Akhavian and Behzadan, (2016), N. Li et al. (2016) | SVC | Liu et al. (2018) |
| LDA | Zhao et al. (2018) | SVM | Kaewwiset et al. (2021), Jain et al. (2020), Fallucchi et al. (2020), Anh et al. (2020), Khera and Divya (2018), |

| | | | Yadav et al. (2018), Zhao et al. (2018), Jebelli et al. (2018), (Akhavian and Behzdan (2016) |
|---|---|---|---|
| LFM-GcForest | Xie (2020) | XGBoost | Singer and Cohen (2020), Zhao et al. (2018) |

Table 4 summarizes the AI models and their applications in the various stages of HRM. In other words, these sixteen models outperformed other models in data related to HRM. RF is a model whose high performance has been confirmed in three different stages: on-boarding (Kaewwiset et al., 2021; Liu, Li, et al., 2019), employability and benefits (Long et al., 2018), and off-boarding (Jain et al., 2020; Liu et al., 2018). In addition, the SVM is another model whose high performance has been proven to address retention (Jebelli et al., 2018) and off-boarding (Khera & Divya, 2018) issues. Another model whose performance has been approved in more than one stage of the EL is the Adaboost model, whose performance has been approved in the stages of employability and benefits (Liu, Wang, et al., 2019) and off-boarding (Liu et al., 2018). DT is also another model that its high performance is proved in two stages of EL, i.e., recruitment (Zaman et al., 2018) and off-boarding (Yadav et al., 2018). The rest of the models have been examined only in one stage of the EL stages, but among them is the ANN model whose performance in the on-boarding stage has been examined and confirmed by two different studies (Akhavian & Behzadan, 2016; Colomo-Palacios et al., 2014). Therefore, a short introduction to these 5 models (i.e., RF, SVM, AdaBoost, ANN, and DT) are provided as follows.

Table 4. The models with the higher performance in human resource management data

| Models | Employee Lifecycle | | | | |
|---|---|---|---|---|---|
| | Recruitment | On-Boarding | Employability and Benefits | Retention | Off-Boarding |
| RF | | ✓ | ✓ | | ✓ |
| SVM | | | | ✓ | ✓ |
| AdaBoost | | | ✓ | | ✓ |
| DT | ✓ | | | | ✓ |
| ANN | | ✓ | | | |
| XGBoost | | | | | ✓ |
| ELANFIS | | | | ✓ | |
| GNB | | | | | ✓ |
| KNN | ✓ | | | | |
| LFM-gcForest | ✓ | | | | |
| MLR | | | | ✓ | |
| MTCNN-MobileNet-LSTM | | ✓ | | | |
| NB | | | ✓ | | |
| Ordinal CART | | | ✓ | | |
| RST | ✓ | | | | |
| LR | | | | | ✓ |

*4.3.1. Random Forest (RF)*
By mixing a set of weak learners to generate a stronger learner, random forests offer an enhancement over the basic decision tree structure (Breiman, 2001). In other words, RF is an ensemble model. To optimize algorithm performance, ensemble techniques use a divide-and-conquer strategy. Random forests are constructed by building a number of decision trees, using bootstrapped training sets and selecting a random sample of m predictors as split candidates from the entire set P predictors for each decision tree.

### 4.3.2. Support Vector Machines (SVM)

SVM is often used as a discriminative classifier to categorize fresh data samples. The fundamental principle of SVM is to design a hyperplane that divides n-dimensional data into two classes and maximize the geometric distance between the nearest data points, referred to as support vectors. Notably, practical linear SVM often produces comparable results to logistic regression (Raschka, 2015).

### 4.3.3. Adaptive Boosting (AdaBoost)

Boosting is a machine learning technique that is based on the concept of combining several very weak and faulty prediction rules to create a highly accurate prediction rule. AdaBoost, an acronym for Adaptive Boosting, is a meta-algorithm for statistical categorization. The output of the other learning algorithms (referred to as 'weak learners') is merged into a weighted sum that reflects the boosted classifier's final output. AdaBoost is adaptive in the sense that it adjusts succeeding weak learners in favor of cases misclassified by prior classifiers. It may be less prone to overfitting than other learning algorithms in neural certain cases (Schapire, 2013).

### 4.3.4. Decision Tree (DT)

The decision tree approach is a supervised technique that uses a tree-like structure to construct classification or regression models. DT is very powerful (Friedman, Hastie, & Tibshirani, 2001) and it can manage missing values and mixed features (Efron & Hastie, 2016), and it is capable of automatically selecting variables (Efron & Hastie, 2016).

### 4.3.5. Artificial Neural Network (ANN)

Neural networks, also known as multi-layer perceptron, are used to imitate the human nervous system's processes. A neural network in its simplest form is a single perceptron. The input values, associated weights, bias, activation functions, and calculated output are all required components of a perceptron. To solve difficult issues, a neural network may comprise several layers between the input and output. Given sufficient hidden units, networks are a universal approximation technique capable of modeling any smooth function to any desired degree of accuracy (Murphy, 2012).

## 5. Conclusions

The integration of technology with HRM has provided the basis for the use of AI models in the management of EL processes. In different stages of HRM, a lot of data is generated that data-driven decision-making approaches use this data to optimize all stages of the EL, from the recruitment stage to the off-boarding stage. It was found that employee attrition and the management of off-boarding stage is a stage on which more research has been done. This indicates the value that employees, which are the intellectual capital of the organization, have for an organization. Most of the articles published at this stage of the EL focus on predicting the factors influencing employees' decision to leave the organization. Intellectual capital belongs to the employees and with their departure from the organization, the organization loses this capital, while it has spent a lot of money on the development of HR. Hence, the loss of manpower is very costly for organizations, especially if the costs of recruiting and training new staff are added to it. However, less research has been done on the retention phase, which requires more attention from researchers.

The findings of this study also revealed that not only machine learning models (such as SVM) and Deep Learning models (such as ANN) have contributed to HRM decision-making, ensemble models (such as RF) and hybrid models (such as Adaboost) have also been developed to address HRM problems. Ensemble models are models that combine two or more machine learning models to increase the predictive power, while hybrid models combine an optimization model with a machine learning or deep learning machine model to increase the accuracy of the model.

Findings of the present study contribute to the literature of HRM and AI using a systematic review of the literature and by providing the state-of-the-art of advancements of AI models in EL management. Organizations, especially HR managers, can use the findings of this study to easily select the appropriate AI model to address the challenges of each stage of the EL and use the benefits and high performance of these models in their decisions. On the other hand, this study provides a foundation for future research. Within the HRM practices, 'retention' and 'off-boarding' represent two extreme challenges for HR managers. Searching new models of 'work' and 'employment' or the 'new-normal' of work and employment in the post-Covid pandemic period, there is a growing need to better understand the 'practice/process of the new model of work (i.e., remote work, telework, etc.). The necessity of new models of work does vary importantly by sectors of economic activities, reflecting the importance of 'physical proximity' syndrome. In addition, the mainstream of HRM literature focuses almost exclusively on the 'Standard Employment Relations' (SER) practice but neglects the fast-growing share of the Non-Standard Employment Relations (Non-SER) - this is the so-called 'prevarication of work' (e.g., grate majority of the web-based platform work mentioned in the conclusion too - practiced in the form of non-SER or in an entrepreneurial status). Therefore, the use of AI models to address these challenges is also suggested for future research.

Singh, V. K., Singh, P., Karmakar, M., Leta, J., & Mayr, P. (2021). The journal coverage of Web of Science, Scopus and Dimensions: A comparative analysis. *Scientometrics, 126*(6), 5113-5142. https://doi.org/10.1007/s11192-021-03948-5

Susmita, E. K. K. A., & Singh, P. (2022). Predicting HR Professionals' Adoption of HR Analytics: An Extension of UTAUT Model. *Organizacija*, 55(1), 77-93. https://doi.org/10.2478/orga-2022-0006

Xie, Q. (2020). Machine learning in human resource system of intelligent manufacturing industry. *Enterprise Information Systems*, *16* (2), 264-284. https://doi.org/10.1080/17517575.2019.1710862

Yadav, S., Jain, A., & Singh, D. (2018). *Early prediction of employee attrition using data mining techniques*. Paper presented at the 2018 IEEE 8th International Advance Computing Conference (IACC). https://doi.org/10.1109/IADCC.2018.8692137

Zaman, E. A. K., Kamal, A. F. A., Mohamed, A., Ahmad, A., & Zamri, R. A. Z. R. M. (2018). *Staff Employment Platform (StEP) Using Job Profiling Analytics*. Paper presented at the International Conference on Soft Computing in Data Science. https://doi.org/10.1007/978-981-13-3441-2_30

Zhao, Y., Hryniewicki, M. K., Cheng, F., Fu, B., & Zhu, X. (2018). *Employee turnover prediction with machine learning: A reliable approach*. Paper presented at the Proceedings of SAI intelligent systems conference. https://doi.org/10.1007/978-3-030-01057-7_56

Zhe, I. T. Y., & Keikhosrokiani, P. (2021). Knowledge workers mental workload prediction using optimised ELANFIS. *Applied Intelligence, 51*(4), 2406-2430. https://doi.org/10.1007/s10489-020-01928-5
**Appendix**

Table A1. A summary of reviewed articles in this study.

| Authors | Year | Source title |
|---|---|---|
| Li X., Chi H.-L., Lu W., Xue F., Zeng J., Li C.Z. | 2021 | Automation in Construction |
| Teoh Yi Zhe I., Keikhosrokiani P. | 2021 | Applied Intelligence |
| Kaewwiset T., Temdee P., Yooyativong T. | 2021 | 2021 Joint 6th International Conference on Digital Arts, Media and Technology with 4th ECTI Northern Section Conference on Electrical, Electronics, Computer and Telecommunication Engineering, ECTI DAMT and NCON 2021 |
| Singer G., Cohen I. | 2020 | Entropy |
| Chuang Y.-C., Hu S.-K., Liou J.J.H., Tzeng G.-H. | 2020 | Technological and Economic Development of Economy |
| Jain P.K., Jain M., Pamula R. | 2020 | SN Applied Sciences |
| Xie Q. | 2020 | Enterprise Information Systems |
| Fallucchi F., Coladangelo M., Giuliano R., De Luca E.W. | 2020 | Computers |
| Anh N.T.N., Tu N.D., Solanki V.K., Giang N.L., Thu V.H., Son L.N., Loc N.D., Nam V.T. | 2020 | International Journal of Sensors, Wireless Communications and Control |

| Authors | Year | Source |
|---|---|---|
| Jayadi R., Firmantyo H.M., Dzaka M.T.J., Suaidy M.F., Putra A.M. | 2019 | International Journal of Advanced Trends in Computer Science and Engineering |
| Liu J., Wang T., Li J., Huang J., Yao F., He R. | 2019 | Conference Proceedings - IEEE International Conference on Systems, Man and Cybernetics |
| Liu J., Li J., Wang T., He R. | 2019 | Proceedings - 5th IEEE International Conference on Big Data Service and Applications, BigDataService 2019, Workshop on Big Data in Water Resources, Environment, and Hydraulic Engineering and Workshop on Medical, Healthcare, Using Big Data Technologies |
| Khera S.N., Divya | 2019 | Vision |
| Kamaru Zaman E.A., Ahmad Kamal A.F., Mohamed A., Ahmad A., Raja Mohd Zamri R.A.Z. | 2019 | Communications in Computer and Information Science |
| Moyo S., Doan T.N., Yun J.A., Tshuma N. | 2018 | Human Resources for Health |
| Yadav S., Jain A., Singh D. | 2018 | Proceedings of the 8th International Advance Computing Conference, IACC 2018 |
| Long Y., Liu J., Fang M., Wang T., Jiang W. | 2018 | ACM International Conference Proceeding Series |
| Liu J., Long Y., Fang M., He R., Wang T., Chen G. | 2018 | ACM International Conference Proceeding Series |
| Zhao Y., Hryniewicki M.K., Cheng F., Fu B., Zhu X. | 2018 | Advances in Intelligent Systems and Computing |
| Jebelli H., Khalili M.M., Hwang S., Lee S. | 2018 | Construction Research Congress 2018: Safety and Disaster Management - Selected Papers from the Construction Research Congress 2018 |
| Akhavian R., Behzadan A.H. | 2016 | Automation in Construction |
| Li N., Kong H., Ma Y., Gong G., Huai W. | 2016 | International Journal of Advanced Manufacturing Technology |
| Colomo-Palacios R., González-Carrasco I., López-Cuadrado J.L., Trigo A., Varajao J.E. | 2014 | Information Systems Frontiers |

Table A2. Acronyms.

| Acronym | Explanation |
|---|---|
| AdaBoost | Adaptive Boosting |
| ANFIS | adaptive neuro-fuzzy inference system |
| AI | Artificial Intelligence |
| ANN | Artificial Neural Network |
| DEMATEL | Decision Making Trial and Evaluation Laboratory |
| DT | Decision Tree |
| EL | Employee Lifecycle |
| XGBoost | Extreme Gradient Boosting |
| ELANFIS | Extreme Learning Adaptive Neuro-Fuzzy Inference System |
| Fed | Federated Learning |
| GRU | Gated Recurrent Unit Neural Network Framework |
| GNB | Gaussian Naïve Bayes |
| GBT | Gradient Boosting Trees |
| HR | Human Resource |
| HRM | Human Resource Management |
| KNN | K-Nearest Neighbors |
| LFM | Latent Factor Model |
| LDS | Linear Discriminant Analysis |
| LSVM | Linear Support Vector Machines |
| LR | Logistic Regression |
| LSTM | Long Short-Term Memory |
| gcForest | Grained Cascade Forest |
| MLP | Multi-Layer Perceptron |
| MLR | Multinomial Logistic Regression |
| MADM | Multiple-Attribute Decision-Making |
| MTCNN | Multi-Task Cascaded Convolutional Networks |
| MB | Multivariate Bernoulli |
| NB | Naïve Bayes |
| Ordinal CART | Ordinal Classification and Regression Tree |
| PRISMA | Preferred Reporting Items for Systematic Reviews and Meta-Analyses |
| RF | Random Forest |
| ROS | Random Over-Sampling |
| RST | Rough Set Theory |
| SER | Standard Employment Relations |
| SVC | Support Vector Classifier |
| SVM | Support Vector Machine |
| SWP | Smart Work Packaging |

Authors' Biography

Dr. Saeed Nosratabadi has recently graduated from the Doctoral School of Economic and Regional Sciences, Hungarian University of Agriculture and Life Sciences, Godollo, Hungary. His main research interests are digital transformation, sustainability, and business models.

Roya Khayer Zahed holds a PhD in Business Management in the Faculty of Administrative Sciences and Economics at the University of Isfahan in Iran. She has taught management courses in the past eight years and is an employee in Iranian National Tax Administration, Fars province. Her areas of interest include human resource management, talent management and organizational behavior. She has several articles in these areas.

Vadim V. Ponkratov, Ph.D. in Economics, director of Financial Policy Center of the Public Finance Department in Financial University under the Government of the Russian Federation; corresponding member of International Academy of Technological Sciences. Obtained PhD degree in 2006 in Financial Academy under the Government of the Russian Federation. Member of Government and Federal Assembly of the Russian Federation working groups; member of energetic strategy and developing fuel-energetic complex committee in the chamber of commerce and industry of the Russian Federation; member of expert council for improving tax legislation and law enforcement practice in the chamber of commerce and industry of the

Russian Federation. Published more than 150 works. Area of scientific interest: government financial policy and assessing its effectiveness; managing oil and gas budget income, strategies of forming and implementing means of sovereign funds; natural rent and instruments for its expropriation; instruments of government financial stimulation of economic development. https://orcid.org/0000-0001-7706-5011

Evgeniy V. Kostyrin - professor of the department of Finances in Bauman Moscow State Technical University. Graduated as Doctor of Economic sciences in Bauman Moscow State Technical University. He is also cooperating with Moscow State University of Medicine and Dentistry. Prof. Kostyrin's research interests are models of medical services administration, economical-mathematical modeling processes of medical organizations development administrations, medical savings accounts as advanced technology of Russian Federation healthcare funding, sovereign issue as an instrument of salary and Russian economy growth. https://orcid.org/0000-0003-2569-1146